\documentclass[fleqn,usenatbib]{mnras}

\usepackage{newtxtext,newtxmath}

\usepackage[T1]{fontenc}

\DeclareRobustCommand{\VAN}[3]{#2}
\let\VANthebibliography\thebibliography
\def\thebibliography{\DeclareRobustCommand{\VAN}[3]{##3}\VANthebibliography}

\usepackage{graphicx}	
\usepackage{amsmath}	

\title[Orbital Stability of 3 and 4-Body Systems]{Orbital Stability of Hierarchical 3 and 4-Body Systems with Inclination: \\
Results for Kepler-1625, 1708, and HD 23079}

\author[S. D. Patel et al.]{
Shaan D. Patel,$^{1}$\thanks{E-mail: shaan.patel@uta.edu}
Billy Quarles,$^{2}$\thanks{E-mail: billylquarles@gmail.com}
Manfred Cuntz$^{1}$\thanks{E-mail: cuntz@uta.edu}
\\
$^{1}$Department of Physics, University of Texas at Arlington, Arlington, TX 76019, USA\\
$^{2}$Department of Physics and Astronomy, East Texas A\&M University, Commerce, TX 75428, USA\\
}

\date{Accepted XXX. Received YYY; in original form ZZZ}

\pubyear{2025}

\begin{document}
\label{firstpage}
\pagerange{\pageref{firstpage}--\pageref{lastpage}}
\maketitle

\begin{abstract}
As the number of potential exomoon candidates grows, there is a heightened motivation of pursing orbital stability analyses. In this work, we provide an in-depth investigation into 4-body systems, consisting of a star, planet, moon, and submoon by using the N-body simulator \texttt{rebound}.  Particularly, we focus on the system of Kepler-1625, where evidence of a possible exomoon has been obtained.  We investigate the 3-body star--planet--moon system for the proposed exomoon parameters allowing us to identify stable regions associated with most of the space parameters. Thereafter, we consider a 4-body system including a potential submoon. We find that there are both stable and unstable regions, as expected, as well as resonance patterns that are further explored using numerical and analytical methods including secular perturbation theory.  We are able to identify these resonances as secular in nature. In addition, we investigate 3-body versions of two other systems, Kepler-1708 and HD 23079, while also studying a 4-body version of HD 23079.  Our work may serve as a generalized framework for exploring other planet--moon cases in the future while noting that the current 4-body study may be an incentive for studying further exomoon and submoon systems.

\end{abstract}

\begin{keywords}
exoplanets -- planets and satellites: dynamical evolution and stability
\end{keywords}



\section{Introduction}

Exoplanets and exomoons are considered pivotal topics of contemporaneous astronomy.  More than 5000 exoplanets have been detected to date, mostly through the radial velocity and transit methods --- see the NASA Exoplanet Archive\footnote{\url{https://exoplanetarchive.ipac.caltech.edu/}}; \cite{Akeson2013}. Confirmation of exomoon candidates remains elusive, where surveys are biased towards short period planets and against widely separated satellites. Nonetheless, the existence of exomoons is strongly implied based on the statistics of the Solar System objects\footnote{\url{https://ssd.jpl.nasa.gov/}} noting that the number of moons exceeds the number of planets by a factor of 20 or more, depending on whether Solar System moons of minuscule size are considered or not.

Targeted efforts of finding exoplanets have dominated astronomy and astrophysics for many years, an activity which is now supplemented by the search for exomoons.  As the number of exomoon candidates continues to grow, there is also an emerging need for theoretical studies to help interpret these candidates.  Examples of previous work include contributions by \citet{1997Natur.385..234W}, \citet{2013AsBio..13...18H}, \citet{2014AsBio..14..798H}, and \citet{2017MNRAS.472....8Z} as these studies focus on a large range of topics relevant to exomoon-based exolife, as well as on orbital stability, tidal heating, and climate modeling.  Analyses on the expected survival of exomoons in different dynamical environments have been given by \citet{Dobos_2021}.

There have been multiple attempts to identify exomoons with the two best candidates coming from the Kepler Mission, Kepler-1625 and Kepler-1708.  Kepler-1625 is a solar-type star (or perhaps a subgiant), with an effective temperature of about 5550~K and a radius 1.7 times larger than the Sun; see \citet{Teachey2018} for further information.  Its companion has a mass less than ${\sim}11.6$ Jupiter masses (\citealt{2020A&A...635A..59T}) with the planetary eccentricity readily assumed as zero.  Kepler-1708 consists of a spectral type F star \citep{Stassun2019} that is slightly larger $({\sim}1.12\ R_\odot)$ and more massive $({\sim}1.09\ M_\odot)$ than the Sun, which makes it also hotter $({\sim}6100\ {\rm K})$ and more luminous ${\sim}1.5\ L_\odot)$, where $R_\odot$, $M_\odot$, and $L_\odot$ represent solar values; see \cite{Kipping2022}.  It hosts a gas giant slightly smaller than Jupiter, which was not identified as a candidate planet until 2019 (see \citet{2019AJ....157..248H} and \citet{2019AJ....157..218K}) and confirmed in \cite{Kipping2022}.

Both exomoon candidates have courted some controversy, where \cite{Kreidberg2019} and \cite{Heller2019} contested the exomoon interpretation from the Kepler data for Kepler-1625.  In addition, \cite{Heller2024} argued that large exomoons were unlikely in both Kepler-1625 and Kepler-1708.  \cite{Teachey2020} addressed the issue raised by \cite{Kreidberg2019} and \cite{Heller2019} more systematically, while \cite{Kipping2024} showed that the analysis method (as presented) by \cite{Heller2024} actually strengthened the exomoon interpretation of the data.  With rigorous debate and an openness concerning the data analysis, Kepler-1625b-i and Kepler-1708b-i remain the two best exomoon candidates to date.

\newpage


Using the Kepler-1625 system, \citet{Rosario-Franco2020} revised the general stability limit for prograde satellites to 40\% of the host planet's Hill radius $R_{\rm H}$ using a more representative initial mean anomaly.  On the other hand, \cite{Domingos2006} found the maximum stability limit at ${\sim}0.5R_{\rm H}$ (assuming a host planet in a circular orbit) with $R_{\rm H}$ given as
\begin{equation} \label{hill_radius}
    R_{\rm H} \approx a(1-e)\sqrt[3]{\frac{m}{3M}} \ ,
\end{equation}
where $a$ and $e$ are the orbital elements of the smaller body with mass $m$ orbiting a larger body with mass $M$.

\citet{Rosario-Franco2020} also investigated the principal possibility of submoons, i.e., moons of moons. No submoons currently exist in the Solar System, while the previous possibility of submoons in the Solar System is spurious at best\footnote{It has been argued that Saturn's satellite Iapetus possessed a submoon in the distant past, which is one of several proposed hypotheses to account for Iapetus's unusual equatorial ridge (\citealt{Levison_2011}, \citealt{Dombard_2012}).}.  Submoons may still exist as one more layer in the star--planet--moon hierarchy, in consideration of the vast diversity of arrangements for planetary and/or stellar bodies as currently identified.

In a subsequent study, \citet{Jagtap_2021} re-evaluated the outer edge of orbital stability for possible exomoons in HD 23079.  Here a solar-type star is host to a Jupiter-mass planet in a nearly circular orbit in the outer stellar habitable zone; see \citet{2002ApJ...571..528T} for details.  \citet{Jagtap_2021} deduced the outer exomoon stability limit using both N-body and tidal migration simulations considering a large range of initial conditions, encompassing both prograde and retrograde orbits. While current technologies are incapable of detecting exomoons in the HD 23079 system, the authors did comment on the detectability of putative moons through Doppler monitoring within direct imaging observations using the future research capacities of extremely large telescopes.  Previous results for exomoons in the HD 23079 systems have been obtained by \citet{Cuntz2013}.  

The purpose of this study is to provide a framework for exploring exomoon candidates and to investigate the general possibility of submoons. Through this framework, we intend to assist in the interpretation of observational exomoon data, which could expand the dynamics of the system by including a potential submoon and explore the stability of this fourth body.  Though our main focus is on Kepler-1625, thereafter, we apply our framework to two other cases: Kepler-1708 and HD~23079.  For Kepler-1708, the parameters are based on \citet{Kipping2022}, and for HD~23079, they are from \citet{2020MNRAS.492..377W} and \citet{2015AA...575A..18B}.  While we investigate Kepler-1625 and HD~23079 both as 3 and 4-body systems, we choose to study Kepler-1708 only as a 3-body system considering that the proposed exomoon being very close to the planet, which would make a 4-body simulation computationally very expensive.

In Section 2, we discuss our \texttt{rebound} simulation setup and initial conditions along with the numerical and analytical methods adopted for analyzing our parameter space results.  Section 3 shows our main results for our 3-body and 4-body simulations and analyses, whereas Section 4 conveys our summary and conclusion.

\section{Methods}

\subsection{\texttt{rebound}}

We use the N-body software \texttt{rebound} \texttt{(v3.25.1)} \citep{rebound} to pursue simulations that explore the orbital stability of exomoons and submoons. The quality of N-body simulations depend on a number of timescales set by the user, including the total simulation time, the integration timestep, and the output frequency.  For our simulations, a timescale of 10$^5$ years is sufficient for determining the stability of moons and submoons because the orbital periods are much less than 1 year \citep{Rosario-Franco2020}. We use an integration timestep that is a fraction of the initial orbital period for the submoon (2\%) in the 4-body simulations and for the moon (5\%) in the 3-body simulations to mitigate potential sources of numerical error in the orbital integrations. 

A sufficiently small integration timestep helps mitigate error along with an appropriate integration algorithm.  These types of simulations typically use the \texttt{WHFast} integrator \citep{reboundwhfast} along with a corrector of 11th order to balance wall time and error minimization. We run test simulations with the \texttt{WHFast} integrator, where we found results inconsistent with known results involving the von
Zeipel--Lidov--Kozai mechanism\citep{vonZeipel1910,Lidov1962,Kozai1962}  in the 3-body problem and were led to test other integration methods in \texttt{rebound}.  

We use 9 representative points in a parameter space orbital inclination and semimajor axis of a given satellite (moon or submoon), where we run simulations with those points using 4 different integrators: \texttt{WHFast}, \texttt{WHCKL}, \texttt{SABA(10,6,4)} \citep[and ref. therein]{saba,reboundhighorder}, and \texttt{IAS15} \citep{reboundias15}.  The \texttt{IAS15} integrator is our base case because it applies an adaptive timestep to maintain the numerical error near machine precision, albeit with a much longer wall time.We find that \texttt{SABA(10,6,4)} provides results similar to \texttt{IAS15}, but in a shorter wall time. The remainder of our simulations uses the \texttt{SABA(10,6,4)} integrator for both the 3-body and 4-body cases.  For an in-depth discussion concerning the differences between integration methods, see \cite{reboundhighorder}.

Even with the \texttt{SABA(10,6,4)} integrator, the simulations with a high initial inclination for the submoon did not give a sufficient maximum eccentricity for the submoon as expected from the hierarchical 3-body secular approximation \citep{Naoz2016} that is applicable to the planet--moon--submoon case.  Such a discrepancy indicates an issue with coordinate frames when extracting the orbital elements from our \texttt{rebound} simulations.

The expected maximum eccentricity values are given as
\begin{align} \label{max_e}
    e_{\rm max} &= \sqrt{1-\frac{5}{3}\cos^{2}i_{\rm o}}
\end{align}
with $i_{\rm o}$ denoting the initial inclination.  When initializing the particles, \texttt{rebound} \texttt{(v3.25.1)} allows for the transformation of a single particle hierarchy (i.e., star--planet--moon).  After inspecting the initial coordinates provided by \texttt{rebound} in our 4-body case, we found that a second transformation was not applied to incline the orbit of the submoon relative to the host moon.  As a result, we apply coordinate rotations for both the moon and the submoon to ensure the correct transformation from hierarchical orbital elements to Cartesian coordinates.  Following this step, we convert the Cartesian coordinates to Jacobi coordinates to confirm whether the outcome by \texttt{rebound} is as expected. We also need to perform the inverse transformation when recording the submoon's hierarchical orbital elements relative to the moon, not the star. This inverse transformation is applied as a post processing step at the end of each simulation. The results are found in alignment with theoretical expectations. These rotations and subsequent counter-rotation are given in Appendix~A.

\subsection{Initial Conditions}

\begin{table*}
	\centering
	\caption{Initial conditions used in 3-body (first row of each respective system) and 4-body (second row) \texttt{rebound} simulations for HD~23079 \citep{2015AA...575A..18B,2020MNRAS.492..377W}, Kepler-1625 \citep{Teachey2018}, and Kepler-1708 \citep{Kipping2022}.  Repeated parameters (i.e., $M_\ast$, $M_{\rm p}$, etc.) are left blank in the 4-body row for HD 23079 and Kepler-1625. All symbols have their usual meaning; see \citet{openstax} for additional definitions. Subscripts for the planet, moon, and submoon are denoted as p, m, and sm, respectively. Planet, moon, and submoon orbital elements are defined with respect to the sun, planet, and moon, respectively.}
	\label{tab:init_con}
	\begin{tabular}{lcccccccccc} 
		\hline
  		 & $M_{\ast}$ & $M_{\rm p}$ & $a_{\rm p}$ & $e_{\rm p}$ & $M_{\rm m}$ & $a_{\rm m}$ & $i_{\rm m}$ & $M_{\rm sm}$ & $a_{\rm sm}$ & $i_{\rm sm}$ \\
             & ($M_{\odot}$) & ($M_{\rm J}$) & (au) & & ($M_{\oplus}$) & (au) & ($^\circ$) & ($M_{\oplus}$) & ($R_{\rm H}$) & ($^\circ$)  \\
            \hline
        HD 23079 & 1.01 & 2.41 & 1.586 & 0.087 & 1.000 & 0.0260--0.0530 & 0--60 & ... & ... & ... \\
        & & & & & & 0.0298 & 9 & $1.67\times10^{-4}$ & 0.20--0.33 & 0--60 \\
		Kepler-1625 & 1.04 & 2.28 & 0.863 & 0 & 10.218 & 0.0170--0.0320 & 0--60 & ... & ... & ... \\
          & & & & & & 0.0219 & 25 & $1.67\times10^{-4}$ & 0.20--0.33 & 0--60 \\
        Kepler-1708 & 1.088 & 1.00 & 1.640 & 0.40 & 17.148 & 0.0038--0.0070 & 0--60 & ... & ... & ... \\
		\hline
	\end{tabular}

\end{table*}

Our simulations explore the orbital stability for two main cases: 3-body systems focusing on the exomoon's stability and 4-body systems focusing on the potential submoon's stability. Each simulation begins with all the bodies added hierarchically, or relative to the center-of-mass for each subsystem. The 3-body exomoon simulations vary the moon's inclination ($0^\circ-60^\circ$, with $1.2^\circ$ steps) and 51 values for the semi-major axes; see Table \ref{tab:init_con} for information on the initial conditions. The parameter ranges are related to the values and error bars given in the literature for the specific system \citep{Teachey2018,Kipping2022}, except HD 23079, where no parameters are available in the literature. In addition to the observational constraints in the literature, the initial inclination range is also motivated by including the range where the von Zeipel--Lidov--Kozai effect is relevant.

For Kepler-1625, the minimum/maximum semi-major axes chosen for the exomoon are the min/max of the 2 sigma error bars from \citet{Teachey2018}. The min/max semi-major axis values for Kepler-1708 are calculated from the error bars given in \citet{Kipping2022}. For HD~23079, there are no proposed exomoon parameters; hence, we choose 20\% to 40\% of the Hill radius as the min/max semi-major axis values based on the general stability limits for exomoons proposed by \citet{Rosario-Franco2020}. Each simulation uses a range of 10 random initial mean anomaly ${\rm MA}$ values to account for any potential bias in stability due to the choice of initial orbital phase for the moon or submoon.

The maximum eccentricity of the moon or submoon is tracked for a given initial condition, where we report the median value over the 10 simulations (with different ${\rm MA}$ values) as one maximal eccentricity value, ${\rm max}\ e$. Since we do not check for collisions, the distance of the moon or submoon from its respective host body is tracked to determine when to stop a simulation prior to the full simulation time.  When this distance exceeds the corresponding Hill radius, the simulation is terminated and categorized as unstable.

Once the exomoon simulations have finished, we choose one specific stable exomoon case based on proposed parameters from the literature to run further simulations on potential submoons; see Table \ref{tab:init_con} for information on the initial conditions.  Our parameter space for the submoon consists of the same 51, 0\textdegree \ to 60\textdegree \ initial inclinations; furthermore, the semi-major axis range is chosen to be between 20\% and 33\% of the moon's Hill radius. This stability range is taken from studies given by \citet{Rosario-Franco2020}. Figure \ref{fig:orbit} depicts a diagram of the initial setup of our 4-body simulations for Kepler-1625 --- it also indicates how the other systems are initialized. 

\subsection{Secular Resonance} \label{sec:num_analysis}



Secular resonances occur when the precession rates (apsidal $\dot{\omega}$ and nodal $\dot{\Omega}$) of two orbital planes overlap.  The two planes can move together through a linear combination of precession rates that allows for a resonant angle between the orbital planes.  To maintain the resonant angle, the orbits can rotate or change shape, where the latter is more important for long-term stability.  The excitation of eccentricity affects the shape of the orbit, where higher eccentricity can lead to scattering events. A full description of the physics and mathematics for secular resonance is provided in more detail within Chapter 7 of \cite{Murray_Dermott_2000}.

To test the potential for instabilities due to precession occurring between the moon and the submoon, we record the moon and submoon's respective arguments of pericenter over a 100 year period, allowing for data collection over multiple secular periods of the submoon. The apsidal precession rate during a given precession cycle follows an approximately linear trend for the moon and submoon.  As a result, we use the $\texttt{polyfit}$ function from \texttt{numpy} to estimate the slope of the linear trend.  This is done in two ways: (1) we cut off points above and below the main curve such that we can get the slope of the line without the short-period perturbations affecting it, and (2) we take the median value of the curve at each short-period perturbation and find the slope of the median points.

Figure \ref{fig:moonw} shows an example of the short and long-term changes in $\omega_{\rm m}$ between 23 and 26 years, where the slope of the best fit line represents $\dot{\omega}_{\rm m}$. The red dots represent the median points every 0.08 years. The purple line represents the best-fit line determined by \texttt{polyfit} using the median values. 
Once we obtain $\dot{\omega}_{\rm m}$ and $\dot{\omega}_{\rm sm}$ (with subscripts m and sm denoting the moon and submoon, respectively), we find the time derivative of the resonant angle is given as

\begin{equation} \label{phiderv}
\dot{\phi} \ = \ 13\dot{\omega}_{\rm m} - 2\dot{\omega}_{\rm sm}.
\end{equation}

We also use the chaos indicator \texttt{MEGNO} \citep[Mean Exponential Growth of Nearby Orbits;][]{Cincotta1999,Cincotta2000,Cincotta2003} that is included within \texttt{rebound}. The \texttt{MEGNO} value $\langle Y \rangle$ is used to distinguish between periodic, quasi-periodic, and chaotic orbits within a dynamical system, where $\langle Y \rangle \mathbf{\rightarrow} 2$ signifies a (periodic or quasi-periodic) orbit while convergence towards this value depends on the total simulation time . Chaotic orbits are characterized by a \texttt{MEGNO} value $\langle Y \rangle > 2$, which can arise due to dynamical instabilities, orbital forcing of eccentricity, or orbital precession associated with resonances.  

Simulations of this type are typically shorter due to their computational expense, however they are used regularly because they demonstrate the potential for chaotic orbits over relatively short timescales. We employ simulations of 100 years for the Kepler-1625 system to explore how the \texttt{MEGNO} value $\langle Y \rangle$ across the parameter space compares to the maximum eccentricity measured in our longer (100,000 yr) simulations. To highlight the emergence of chaotic orbits, we use the base-10 logarithm of the absolute difference of the \texttt{MEGNO} value $\langle Y \rangle$ from two, or $\log_{10}|\langle Y \rangle - 2|$. Due to the shorter simulation length, the underlying structure is more easily seen in this form.

\subsection{Secular Perturbation Theory}

We compare our simulation results with secular perturbation theory to help analyze a sample case in our parameter space plot for Kepler-1625. Before applying our python-based numerical implementation of the secular perturbation theory to Kepler-1625, we test it using Jupiter, Saturn, and a test object to replicate results found in \citet{Murray_Dermott_2000}. Using the initial conditions for these bodies, we recreate the previous results from \cite{Murray_Dermott_2000} and compare to the \texttt{mathematica} scripts to identify the free/forced eccentricity and inclination values for the system. The underlying secular equations are solved using the vertical and horizontal components of eccentricity and inclination vectors given as



\begin{align}\label{hkpq}
    h &= e\sin(\varpi)\\
    k &= e\cos(\varpi)\\
    p &= i\sin(\Omega)\\
    q &= i\cos(\Omega).
\end{align}

The derivatives of these vector components are expanded using the so-called disturbing function, where the final solution is expressed as a linear combination of $\sin$ and $\cos$ functions.  Further derivation for the secular evolution equations for a general three-body problem can be found in \citet{quarleshkpq}.



Once we confirmed that our scripts work for the test case, we then proceed to a 3-body sub-system of Kepler-1625 involving the planet, moon, and submoon. We set the moon as the host body and the planet and submoon as the perturbers. In addition to adopting the orbital elements of the planet and submoon relative to the moon as the initial conditions, we also include the additional rotation of the submoon's inclination by the moon's inclination similar to our initial coordinate rotation in \texttt{rebound}. After applying the secular theory to our system, we then apply our subsequent counter-rotation as detailed in Appendix~A. This results in an inclination and eccentricity time evolution for the secular evolution that qualitatively matches our \texttt{rebound} simulation over the first 100 years.

After this tutorial case, the next step is to extend this kind of testing to our main 4-body case. The Sun is added as a third perturbing body in the system.  Thus, we use a planet-hosting system including the moon, submoon, and the Sun as the three perturbing bodies.  We apply similar coordinate calculations and counter-rotations as for the 3-body case mentioned above. This provides us with our final secular results to be used to compare to our \texttt{rebound} simulation results. 

\section{Results}
\subsection{Exomoon Stability due to 3-body Interactions} \label{sec:moon_stab}

Using \texttt{rebound}, we investigate the orbital stability of potential exomoons orbiting Kepler-1625b, Kepler-1708b, and HD~23079b over 10$^5$ yr. In these simulations, we vary the initial semi-major axis $a_{\rm m}$ and inclination $i_{\rm m}$ of the exomoon while keeping the star and planet's initial parameters fixed and tracking the maximum eccentricity of the exomoon attained over the simulation. The maximum eccentricity is a proxy for stability because a high value correlates with a high probability of ejection beyond the host planet's Hill radius.  We examine the results of our simulations through filled contour maps (see Figure~\ref{fig:m_param}), where each cell is color-coded by the median value of the logarithm of maximum eccentricity attained using a set of 10 initial conditions ($a_{\rm m},\ i_{\rm m}$) that randomly vary the initial mean anomaly $\rm MA_{\rm m}$ of the exomoon.  The red areas represent regions of lower maximum eccentricity (i.e., more stable regions) while yellow, green, blue, and purple represent increasingly higher maximum eccentricity that corresponds to a higher likelihood of instability (i.e., ejection).  Figure~\ref{fig:m_param} shows these results for all 2601 initial condition pairs ($a_{\rm m},\ i_{\rm m}$) using three systems: (a) Kepler-1625, (b) HD 23079, and (c) Kepler-1708.  

Figures~\ref{fig:m_param}a and \ref{fig:m_param}b show a gradient in the maximum eccentricity starting at ${\gtrsim}45^\circ$. This behavior coincides with the von Zeipel--Lidov--Kozai effect that we expect from hierarchical 3-body systems at these high inclinations. Below $45^\circ$, the exomoons show low maximum eccentricity indicating stability. The maximum eccentricity values reach $0.4$ at ${\sim}0.4\ R_{\rm H,p}$ for an exomoon orbiting Kepler-1625b (Fig. \ref{fig:m_param}a) or HD~23079b (Fig. \ref{fig:m_param}b).  \citet{Rosario-Franco2020} showed similar behavior for the case of Kepler-1625b, but for strictly coplanar orbits. There is another gradient of maximum eccentricity as the exomoon's semimajor axis increases that coincides with the perturbation of the star being more prominent as the exomoon is further from the host planet. At ${\sim}0.4\ R_{\rm H,p}$, we also see the maximum eccentricity decrease with higher inclinations. As the inclination increases leading to an increase in the $z$-component of the force, the kicks on the moon from the star tend to be more perpendicular to the motion of the moon. This results in a lower maximum eccentricity at these higher inclinations at ${\sim}0.4\ R_{\rm H,p}$.

Figure~\ref{fig:m_param}c shows that the maximum eccentricity is more stratified with the moon's inclination, where this is likely due to the proposed semi-major axis of exomoon candidate Kepler-1708b-i that is much closer to the host planet. Using the uncertainty given in \citet{Kipping2022}, we iterate through an exomoon semi-major axis range that lies between ${\sim}6\%-11\%$ of the Hill radius where the planet's gravity largely dominates in the exomoon's tug-of-war with its host star. This leads to an $e_{\rm max}$ between
${\sim}0.003-0.005$ for simulations with an initial $i_{\rm m}$ under the von Zeipel--Lidov--Kozai limit $(<40^\circ)$. 
For more inclined orbits  $(>40^\circ)$, there are oscillations from the von Zeipel--Lidov--Kozai effect that allow for the maximum eccentricity to increase to ${\sim}0.1$ at $40^\circ$ and ${\sim}0.8$ at $60^\circ$.

In Figures~\ref{fig:m_param}a and \ref{fig:m_param}c, we mark (using black lines) the range in possible orbital parameters as currently suggested by the respective observations \citep{Teachey2018,Kipping2022}. We do not make a similar distinction in Fig.~\ref{fig:m_param}b because an exomoon candidate orbiting HD~23079b has not been confirmed by observations, although several theoretical studies have explored the potential for an exomoon \citep[e.g., ][]{Cuntz2013,Jagtap_2021}.   Figure~\ref{fig:m_param}a  shows that most of the observed range for the exomoon falls in regions of low maximum eccentricity. The vertical inclination bar in both Figures~\ref{fig:m_param}a and \ref{fig:m_param}c extends to high inclination, where the von Zeipel--Lidov--Kozai effect can drive the exomoon to higher eccentricity; however, the majority of the bar lies in the nearly circular (more stable) regions.  In both Figures~\ref{fig:m_param}a and\ref{fig:m_param}c, the horizontal semi-major axis bar is always within stable regions. From these visualizations, we conclude that most of the observed parameter ranges are orbitally stable (i.e., bound orbits) except for certain high inclination ranges, i.e., $>40/45^\circ$ for Figures~\ref{fig:m_param}a and \ref{fig:m_param}c, respectively.

\begin{figure*}
	\includegraphics[scale=0.7]{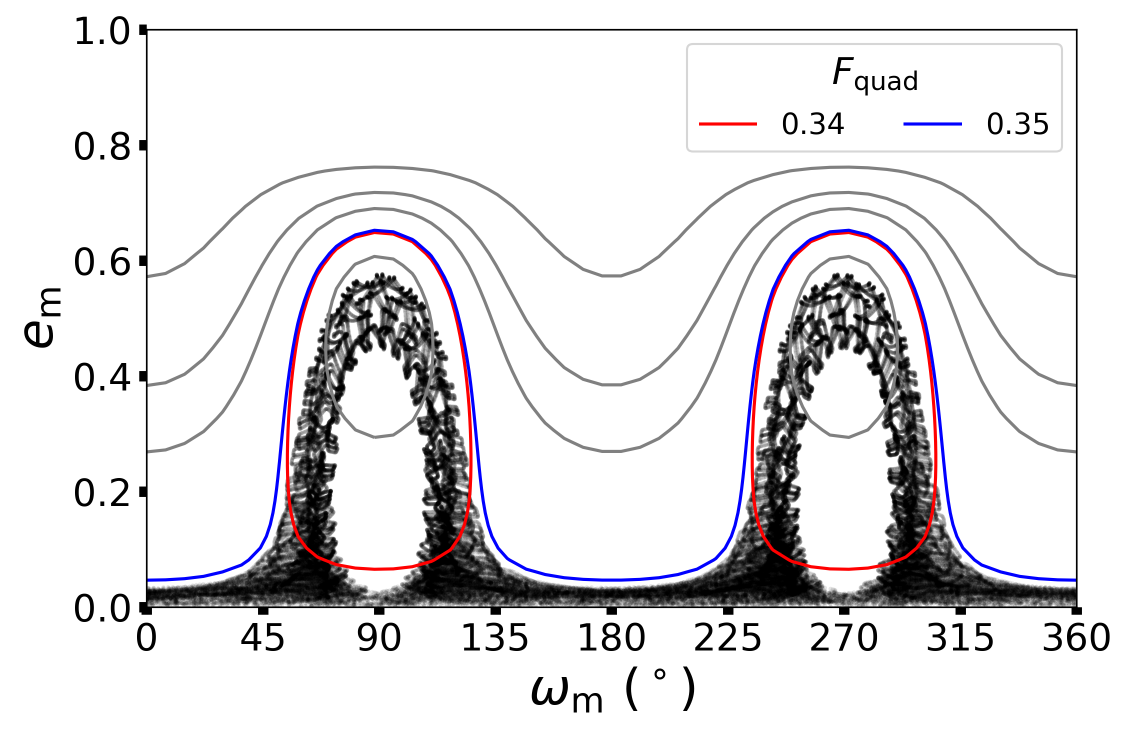}
    \caption{Time evolution of the exomoon's parameters in the $e-\omega$ plane (black dots) through our Kepler-1625 3-body numerical simulations (with $a_{\rm m}=0.0218$~au, $i_{\rm m}=54^\circ$, $\rm MA_{\rm m}=0^\circ$).  The overlaid contours represent the trajectories within the test particle quadrupole approximation; see \citet{Naoz2016}. These general contours for different values of $F_{\rm quad}$ are plotted in gray while the red and blue contours represent the specific trajectories that best match our data.}
    \label{fig:kozai_overlay}
\end{figure*}

To better understand the von Zeipel--Lidov--Kozai effect, we examine a high-inclination case from our 3-body simulations for an exomoon orbiting Kepler-1625b ($a_{\rm m}$: 0.0218~au, $i_{\rm m}$: $54^\circ$). Figure~\ref{fig:kozai_overlay} shows the evolution of the putative exomoon's eccentricity and argument of pericenter for the first 100 years of the simulation. The two distinct libration regions are evident around 90\textdegree \ and 270\textdegree. Overlaid on this plot are two von Zeipel--Lidov--Kozai contours that are calculated from \citet{Naoz2016}. The theoretical von Zeipel--Lidov--Kozai contours do not fully align with the simulated points because the von Zeipel--Lidov--Kozai effect does not take into account the full dynamics of the system (e.g., changing semi-major axis values, among other effects).  Nevertheless, the general form of the von Zeipel--Lidov--Kozai effect is reproduced in Fig. \ref{fig:kozai_overlay}.

\subsection{Orbital Stability of Inclined Submoons} \label{sec:submoon_stab}

\subsubsection{General Approach}
Using our results from Section \ref{sec:moon_stab}, we consider the potential orbital stability of submoons orbiting the Kepler candidate exomoon, Kepler-1625b-i, or a hypothetical exomoon in orbit about HD 23079b.  \citet{Rosario-Franco2020} considered the potential orbital stability for coplanar submoons orbiting Kepler-1625b-i, where our simulations expand the parameter space to also include orbits inclined to the host moon.  

\begin{figure*}
	\includegraphics[scale=0.97]{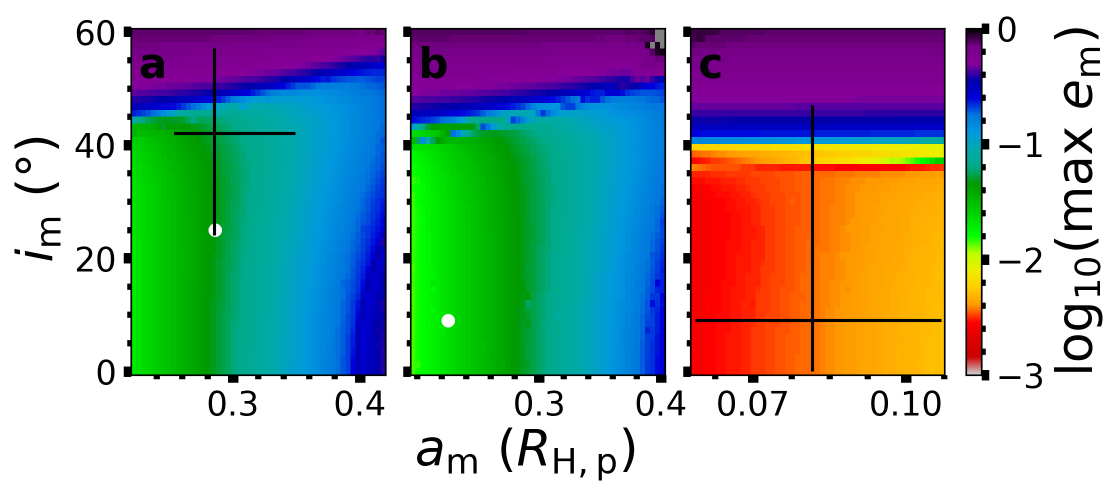}
    \caption{Logarithm of maximum eccentricity ($\rm log_{10}(\text{max}\ {\it e}_m)$; color coded) from stability simulations that vary the exomoon's initial inclination and semi-major axis in (a) Kepler-1625, (b) HD~23079, and (c) Kepler-1708.  The black lines denote the derived parameters for the exomoon given in the literature by (a) \citet{Teachey2018} and (c) \citet{Kipping2022}.  The white dot represents the exomoon parameters chosen for the subsequent 4-body simulations in Section \ref{sec:submoon_stab}.}
    \label{fig:m_param}
\end{figure*}

Akin to our investigation in Section \ref{sec:moon_stab}, we vary the putative submoon's initial semimajor axis and inclination while keeping the initial orbital parameters of the planet and host moon fixed.  The parameters are given in Table \ref{tab:init_con}. The initial conditions for the host moon are also given as white dots in Fig.~\ref{fig:m_param}a and b. For Kepler-1625, we select the lowest inclination that fits within the derived parameters given in the literature.  For HD 23079, we choose an initial condition (somewhat arbitrarily) from Fig. \ref{fig:m_param}b as there are no known exomoon parameters for that system.

The host moon is inclined by $i_{\rm m}$ relative to the planet-star plane (see white dots in Fig. \ref{fig:m_param}), while the submoon's inclination $i_{\rm sm}$ is measured relative to the host moon's inclined orbital plane.  This necessitates applying the respective rotation matrices twice (see Appendix \ref{sec:appendix}) to obtain the correct Cartesian coordinates corresponding to our proposed orbital elements.

Figure \ref{fig:sm_param} illustrates our numerical results using a filled contour map, where each cell is color-coded and represents the logarithm of the maximum eccentricity $({\rm max}\  e_{\rm sm})$ of the putative submoon.  Since the starting orbital phase can influence the overall stability of a body near resonance, we evaluate each set of initial conditions ($a_{\rm sm},\ i_{\rm sm}$) with 10 randomized mean anomalies.  The gray cells represent unstable regions (i.e., $e_{\rm sm}>1.0$), where the submoon is unbound from its host moon.  The resulting structures in the filled contour map highlight the general dynamics of the phase space, which is our focus, and higher-resolution maps may be needed for in-depth analyses that are beyond the scope of our work.

In Fig. \ref{fig:sm_param}, the unstable (gray) region begins at ${\sim}40^\circ$ as expected from the von Zeipel--Lidov--Kozai effect.  The submoon's orbital eccentricity is easily excited to high values ($e_{\rm sm} \gtrsim 0.4$) on a short timescale via the von Zeipel--Lidov--Kozai effect, where the host planet effectively strips the submoon away from the host moon.  This did not occur for the 3-body simulations in Figure \ref{fig:m_param} (see the green and blue regions) because the Hill radius of the host planet is much larger and can accommodate a putative exomoon with a larger eccentricity.  Additionally, there is a difference in timescale where the submoon in the 4-body case experiences many more orbits compared to the exomoon in the 3-body simulations of Fig. \ref{fig:m_param}. The shorter dynamical timescale for the submoon creates additional opportunities for the submoon to leave the moon's Hill radius or become unbound ($e_{\rm sm}>1.0$). 

Figure \ref{fig:sm_param}a shows three distinct (blue-cyan) curves representing areas where the maximum eccentricity value ${\rm max} \ e_{\rm sm}$ ranges between 0.2 and 0.5.  These curves represent regions where secular resonances are possible, which we explore numerically in more detail in Section \ref{sec:secres_num}.  Figure \ref{fig:sm_param}b also shows similar curves but are less distinct and overall less stable (i.e., more gray cells mixed in). HD~23079b begins on an eccentric orbit while Kepler-1625b starts with a circular orbit, which explains the differences in stability between the two features.

\begin{figure*}
	\includegraphics[scale=0.9]{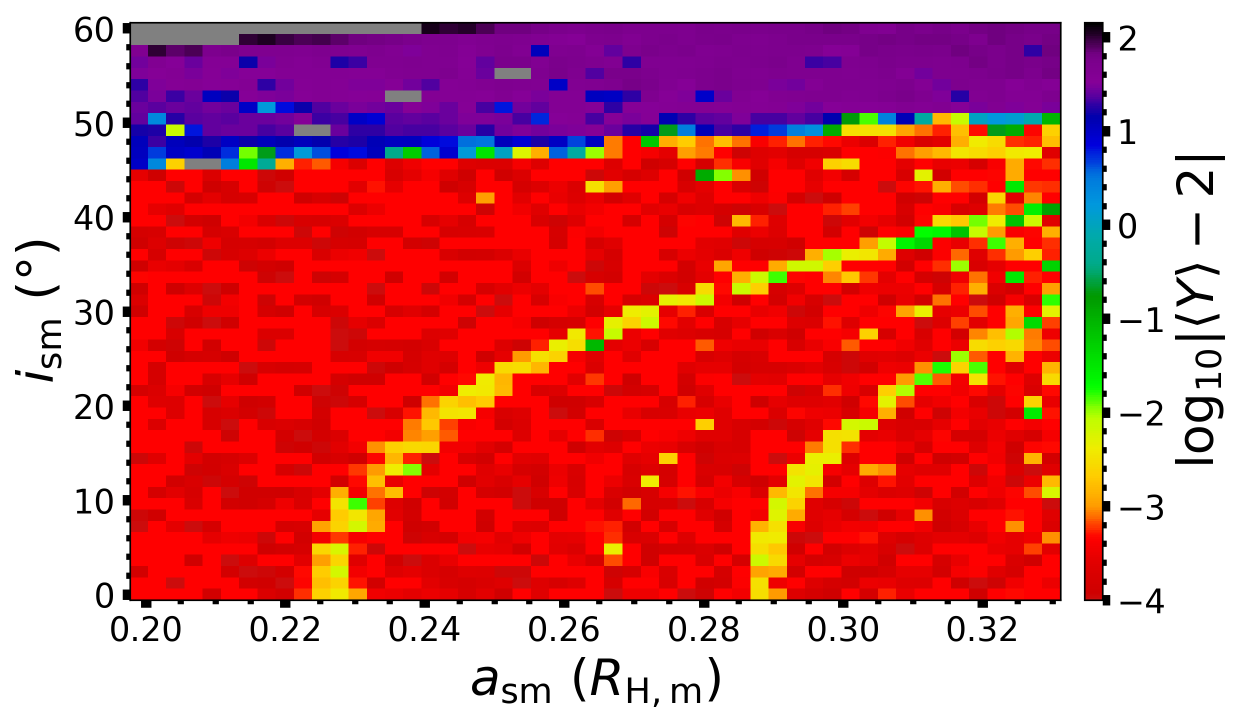}
    \caption{Map of \texttt{MEGNO} values centered around 0 ($\rm log_{10}\vert\text{ \textlangle Y \textrangle}-2\vert$) from 4-body stability simulations for Kepler-162 that vary the submoon's initial inclination and semi-major axis.}
    \label{fig:megno}
\end{figure*}

We use the chaos indicator \texttt{MEGNO} $\langle Y \rangle$ to confirm chaotic or resonant orbits within our parameter space of the submoon's inclination $i_{\rm sm}$ and semimajor axis $a_{\rm sm}$. Figure \ref{fig:megno} shows the results of our 100-year simulations for Kepler-1625b-i where the color-code uses \texttt{MEGNO} $\langle Y \rangle$.  We use the absolute difference of the \texttt{MEGNO} value from two to indicate how far a simulation resides compared to the expected value for regular orbits.  Due to our relatively short runtime, these differences are small and we use the base-10 logarithm to highlight any latent structure.  

The simulations quickly become chaotic for $i_{\rm sm} \gtrsim 45^\circ$, where Fig. \ref{fig:sm_param}a shows that they are unstable as well.  For $40^\circ < i_{\rm sm} < 45^\circ$, these initial conditions are expected to be unstable for longer timescales than we simulated as shown by Fig. \ref{fig:sm_param}a. For $i_{\rm sm}\leq 40^\circ$, we recover the two major structures (in yellow) from Fig. \ref{fig:sm_param}a that begin at $a_{\rm sm} \sim 0.23\ R_{\rm H,m}$ and $a_{\rm sm} \sim 0.29\ R_{\rm H,m}$ for coplanar orbits.  The third curve in Fig. \ref{fig:sm_param}a that begins at $a_{\rm sm}\sim 0.2\ R_{\rm H,m}$ does not appear, where this may appear for longer runtimes.  The appearance of the structures strongly suggests a secular resonance between the submoon-moon and moon-planet orbits.

\subsubsection{Secular Resonance for Submoons} \label{sec:secres_num}

\begin{figure*}
	\includegraphics[scale=0.97]{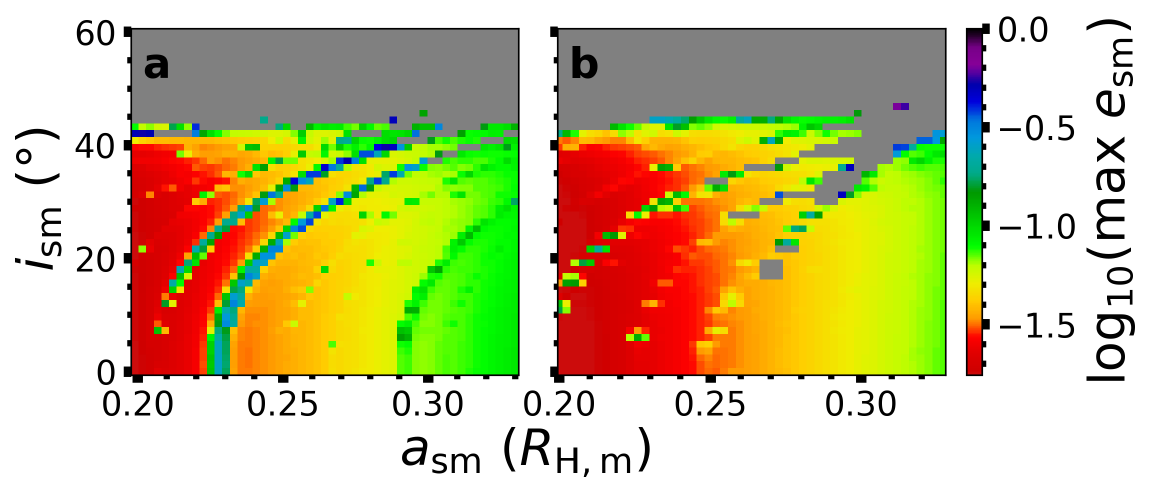}
    \caption{Logarithm of maximum eccentricity ($\rm log_{10}(\text{max}\ {\it e}_{sm})$; color coded) from stability simulations that vary a putative submoon's initial inclination and semi-major axis in (a) Kepler-1625 and (b) HD~23079.  The gray cells denote those initial conditions where the maximum eccentricity of the submoon exceeds $1.0$ and becomes unbound.  The strips of elevated maximum eccentricity $e_{\rm sm}$ are likely due to a secular resonance between the planet--moon and moon--submoon orbital precession rates. }
    \label{fig:sm_param}
\end{figure*}

\begin{figure*}
	\includegraphics[scale=1]{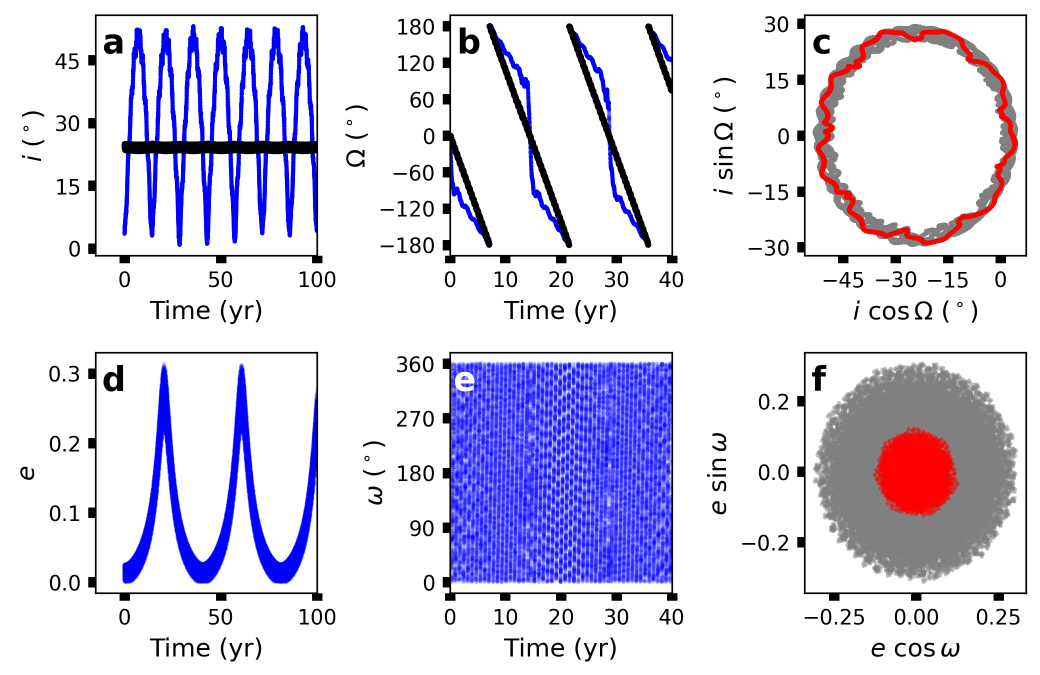}
    \caption{Time series evolution of the moon (black) and submoon (blue) orbital parameters for the 4-body Kepler-1625 simulation with the initial conditions $a=0.00084$~au, $i=3.6^\circ$, and $\rm MA=0^\circ$ \ for the submoon. The star, planet, and moon's initial conditions can be found in Table~\ref{tab:init_con} under Kepler-1625. Each panel depicts the time evolution of the (a) inclination $i$, (b) longitude of ascending node $\Omega$, (d) eccentricity $e$, and (e) argument of pericenter $\omega$. Panel (c) shows the time evolution of two components of the submoon's inclination vector with respect to $\Omega$, while panel (f) depicts two components of the submoon's eccentricity vector with respect to $\omega$. In panels (c) and (f), the red points represent one full cycle while the gray points represent the remainder of the simulation.}
    \label{fig:timeseries}
\end{figure*}

To further explore the previously obtained curves of higher eccentricity apparent in Fig. \ref{fig:sm_param}a, we select one test case on the curve for a more detailed investigation.  We examine the time series evolution for the first 100 years of one case ($a_{\rm sm}=0.00084$~au, $i_{\rm sm}=3.6^\circ$, $\rm MA_{\rm sm}=0^\circ$) in Figure~\ref{fig:timeseries}. The evolution of the submoon's parameters are shown in blue (and red/gray in Figures~\ref{fig:timeseries}c and \ref{fig:timeseries}f) while the moon's parameters are given in black. Though many different orbital elements are depicted, we mostly focus on Figure~\ref{fig:timeseries}a and \ref{fig:timeseries}d, which represent the inclination and eccentricity time evolution, respectively. The oscillations evident in Figures~\ref{fig:timeseries}a and \ref{fig:timeseries}d point to a possible orbital resonance.  The initial inclination and eccentricity are too low to be attributed to the von Zeipel--Lidov--Kozai effect.  

\begin{figure*}
	\includegraphics[scale=0.7]{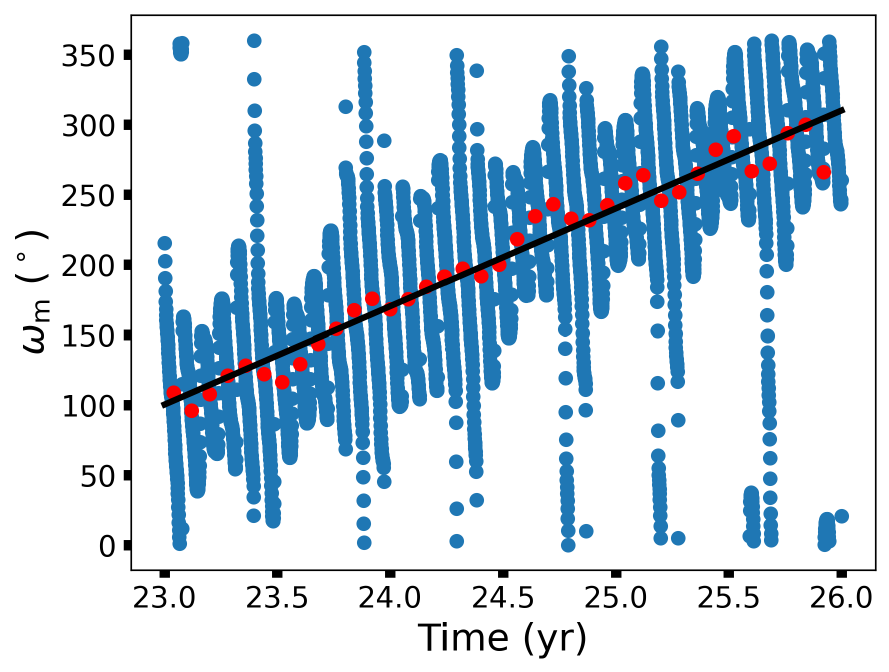}
    \caption{Time series of the moon's argument of pericenter from years 23 through 26 overlaid with the median values taken every 0.08 years. The line of best fit (using \texttt{numpy.polyfit}) for the median points is overlaid.}
    \label{fig:moonw}
\end{figure*}

We investigate the apsidal precession rates of the moon and submoon to check for a secular resonance. As shown in Sect.~\ref{sec:num_analysis}, the apsidal precession rates are determined using linear fits (e.g., with \texttt{numpy.polyfit}) on 100 year simulation data of the arguments of pericenter of the moon and submoon. The 100 year data is broken into smaller sections depending on the period of the trend (i.e., 3 year sections for $\omega_{\rm m}$, see Fig.~\ref{fig:moonw}). In these individual sections, the slope is then calculated by taking the median value over the short-period perturbations, allowing us to determine the slope from these points. From these calculations, we find $\dot{\omega}_{\rm m} = 1.22^\circ\ {\rm yr}^{-1}$ and $\dot{\omega}_{\rm sm} = 8.026^\circ\ {\rm yr}^{-1}$.  The ratio between the apsidal precession rates for the submoon and moon is ${\sim}$13:2, which points to a potential resonance. We find the time derivative of the resonant angle as given in Eq.~\ref{phiderv}. This points to a secular resonance between the precession of these two elements causing an increase in the eccentricity and inclination of the submoon.

\begin{figure*}
	\includegraphics[scale=0.7]{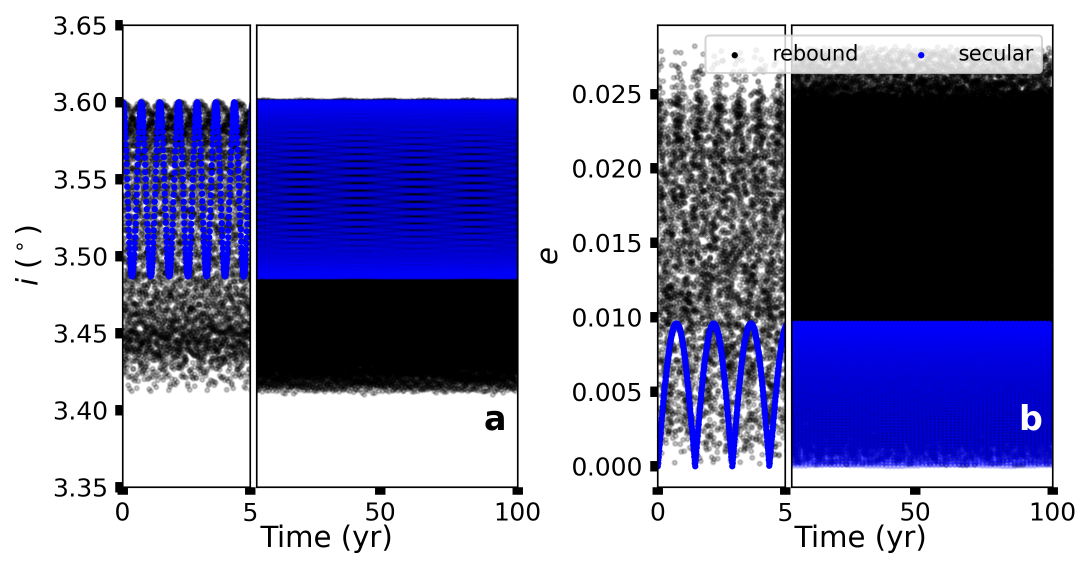}
    \caption{Comparison of the time evolution for the inclination and eccentricity of the submoon in Kepler-1625 using a secular approach (blue) and N-body simulations (black).  Each panel shows the short-term ($<5$ yr) variations on the left, while the longer-term variations are on the right.}
    \label{fig:rebsec_3body}
\end{figure*}

To further explore the potential for secular resonance occurring in our 4-body simulations, we examine a 3-body version of our test case system consisting of a planet--moon--submoon.  By excluding the star, we can compare to previously known results for secular theory among hierarchical triple systems; see Chapter 7 of \citet{Murray_Dermott_2000}. Figure~\ref{fig:rebsec_3body} shows the inclination and eccentricity time evolution of the 3-body system (planet--moon--submoon) using a secular (in blue) and N-body approach using \texttt{rebound} (in black). The moon's inclination and eccentricity are unchanged in both methods, therefore, it is not plotted in Fig.~\ref{fig:rebsec_3body}. 
However, Fig.~\ref{fig:rebsec_3body}a demonstrates that the evolution of submoon's inclination is fairly similar between the secular and N-body approach using \texttt{rebound}, where the N-body simulation results in a larger variation and a slightly lower minimum inclination. Figure \ref{fig:rebsec_3body}b shows that the evolution of the submoon's eccentricity varies more substantially in  the \texttt{rebound} simulation compared to the secular theory approach due to non-secular effects (e.g., mean motion resonances and short-period perturbations) causing additional variations.

\begin{figure*}
	\includegraphics[scale=0.7]{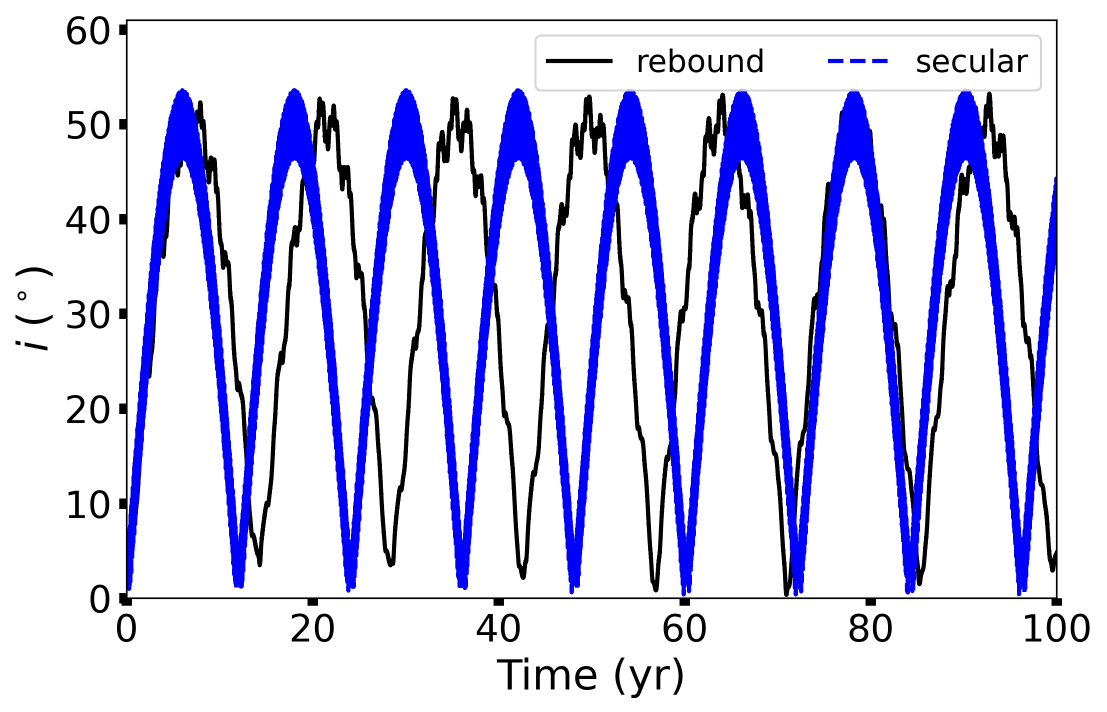}
    \caption{Comparison of the time evolution for the submoon's inclination in Kepler-1625 using a secular approach (blue) and N-body simulations (black).  There is good agreement in the magnitude of the inclination variations, although there are differences in phase.  Note that the maximum inclination is typically more important than the relative phase for determining long-term orbital stability.}
    \label{fig:rebsec}
\end{figure*}

After exploring this trial 3-body subsystem, we then apply this method to our main 4-body case by adding the star as an additional perturber. We compare time evolution of the submoon's inclination with our results from Fig.~\ref{fig:timeseries}a. From this 4-body secular test, we find that the inclination for the submoon matches our \texttt{rebound} case up to a frequency shift. This is detailed in Fig.~\ref{fig:rebsec} where the secular time evolution of inclination for the submoon are overlaid with the inclination from the \texttt{rebound} simulation. As expected, the \texttt{rebound} results have short-term perturbations and a frequency difference caused by non-secular effects; however, we see that the amplitude of the curves are in alignment. This is the second and main test that confirms that the obtained resonance curves are secular in nature. 

%
%
\begin{figure*}
	\includegraphics[scale=0.7]{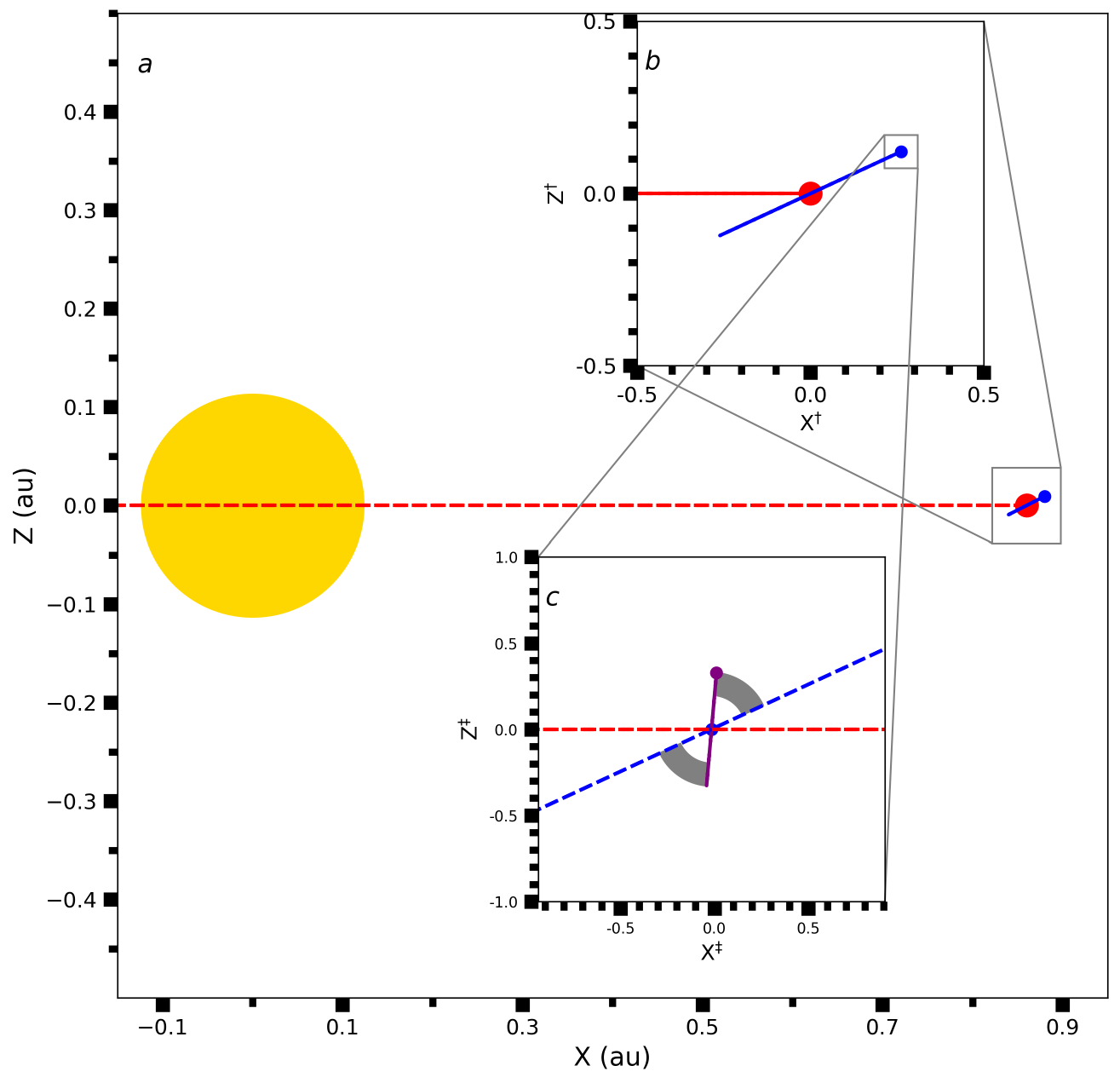}
    \caption{Side view of the 4-body system setup for Kepler-1625 in the \texttt{rebound} simulations: (a) Kepler-1625b (red) orbits Kepler-1625 (yellow) with the proposed exomoon (blue) orbiting Kepler-1625b, (b) the exomoon orbits Kepler-1625b zoomed in showing the inclination of the moon's orbit, and (c) a theoretical submoon (purple) in orbit about the exomoon. The gray areas in panel c represent the parameter space for the submoon adopted in our simulations. The dagger ($\dagger$) and double dagger ($\ddagger$) symbols on the $x$ axis denote that the distance units are in terms of the planet’s Hill radius $R_{\rm H,p}$ and exomoon’s Hill radius $R_{\rm H,m}$, respectively.}
    \label{fig:orbit}
\end{figure*}

%
%

%
%

%
%

%
%

%
%

%
%

%
%

%
%


\section{Summary and Conclusions}

The goal of this study is to provide a framework for exploring exomoon candidates and to investigate the principal possibility of submoons.  Our theoretical framework will assist in the interpretation of current and future observational data by investigating the orbital stability of 3 (star--planet--moon) and 4-body (star--planet--moon--submoon) systems. The resulting stability maps show the orbitally stable configurations of the exomoon or submoon that can exist theoretically which can then be compared to the original observational data. Although there are currently no confirmed exomoon cases, there is strong evidence for their existence.  In this same vein, submoons, i.e., moons of moons, may also exist as one more layer of the star--planet--moon hierarchy.

We focus our simulations on three systems, i.e., Kepler-1625, Kepler-1708, and HD 23079, noting that two of those (i.e., Kepler-1625 and 1708) are the strongest exomoon candidates to date. Exomoon and submoon stability limits in the Kepler-1625 system have been explored previously \citep[e.g., ][]{kollmeier_2018,Rosario-Franco2020,moraes_2022,moraes_2023}.

Although many moons within the Solar System are nearly coplanar relative to the ecliptic, there are notable exceptions including the Moon and Pluto's moon, Charon.  As a result, we numerically simulate 3-body systems (star--planet--moon) for $10^5$ yr where we vary the moon's orbital inclination $i_{\rm m}$ (relative to the planet's orbital plane) and semimajor axis $a_{\rm m}$ (relative to the planetary Hill radius).  For each $(a_{\rm m},\ i_{\rm m})$ pair, we evaluate 10 randomized mean anomalies to incorporate any variations in outcome due to initial orbital phase that could arise from mean motion resonances.  From our simulations (see Fig. \ref{fig:m_param}), we find that the parameter estimates given in \citet{Teachey2018} for the exomoon candidate in the Kepler-1625 system are viable as they mostly fall within the orbitally stable regions (i.e., bound orbits) with a low maximum eccentricity. Higher inclination conditions for the exomoon are predictably more unstable due to the von Zeipel--Lidov--Kozai effect. Similar results are found for Kepler-1708 whose parameters from \citet{Kipping2022} are almost totally situated in the stable regions with even lower maximum eccentricity. Despite the absence of exomoon parameters in the literature, 3-body simulations for HD~23079 show stable regions as well.  They are similar to those of Kepler-1625, which may inform future searches for exomoons in that system.

To evaluate the potential stability of a hypothetical submoon orbiting Kepler-1625b or HD 23079b, we select a stable $(a_{\rm m},\ i_{\rm m})$ pair for a putative exomoon (see the white dots in Fig. \ref{fig:m_param}). Similar to our investigation for 3-body systems, we simulate 4-body systems (star--planet--moon--submoon) that vary the putative submoon's orbital inclination $i_{\rm sm}$ (relative to the host moon's orbital plane) and semimajor axis $a_{\rm sm}$ (relative to the host moon's Hill radius). From these simulations (see Fig. \ref{fig:sm_param}) we identify regions around the putative exomoon in which a submoon could be orbitally stable (i.e., remain bound).  Note that both 4-body systems (Kepler-1625 and HD 23079) have highly unstable regions at $i_{\rm sm} >40^\circ$ due to the von Zeipel--Lidov--Kozai effect, but also have slightly unstable curves that penetrate the stable regions. Furthermore, we explore test cases in the Kepler-1625 4-body system to investigate whether a resonance occurs and what type it could be. From our tests, we find the curve correlates with secular resonance patterns where the misaligned moon and submoon orbits precess together.  We use secular perturbation theory to compare with our \texttt{rebound} simulations and find that the amplitude of the inclination variation in both secular theory and \texttt{rebound} are the same (see Fig.~\ref{fig:rebsec}).

\citet{moraes_2022} discussed the possibility of a second exomoon around Kepler-1625b, assuming the existence of the Neptune-like Kepler-1625b-i. This work shows that a putative second satellite of Kepler-1625b could be orbitally stable if located between Kepler-1625b and Kepler-1625b-i in the coplanar case. \citet{moraes_2023} expands on their prior study by looking at potential co-orbital exomoons in the Kepler-1625 and Kepler-1708 systems. They find that in the tests with 4 bodies (including the star as the central body), the satellites around Kepler-1625b were unstable for most cases. Conversely, the satellites around Kepler-1708b were able to remain stable while being co-orbital.

\citet{Harada_2023} investigated exomoon stability around HIP 41378f using \texttt{rebound} simulations while also varying the inclination and semi-major axis of the exomoon.  They found stability patterns similar to our investigation of 3-body stability, see Fig.~(\ref{fig:m_param}), albeit the von Zeipel--Lidov--Kozai mechanism became relevant at a higher inclination.

Though there is no confirmed submoon cases (yet), a submoon has been proposed to have previously existed in the Solar System around Iapetus, a moon of Saturn, to explain Iapetus's equatorial ridge \citep{Levison_2011,Dombard_2012}. The detection of submoons is primed to be the next stage of discoveries after exomoons; however, as discussed in \citet{Rosario-Franco2020}, this is still a distant thought. Transit methods would require extreme precision for a submoon detection while RV methods would not be possible with current-generation facilities. 

\citet{kollmeier_2018} explored submoon stability using tidal calculations in the Kepler-1625 system. These calculations led to a proposed mass limit of Vesta to Ceres-like submoons; in fact, the submoon masses indicated by those studies are akin to those adopted in our 4-body Kepler-1625 simulations. The authors also noted that submoon inclination were not taken into account, but the submoon inclination would be needed to better understand the dynamics examined in their paper.

Habitability of potential submoons has been studied in \citet{forgan_2018}. This work centers on the climate modeling of a potential Earth-mass submoon in the Kepler-1625 system. The author finds that the submoon as considered could indeed be habitable, although tidal heating from eccentric orbits could affect this outcome.  Although later studies \citep{kollmeier_2018,Rosario-Franco2020} show tidally induced migration would preferentially remove such moons.

Future space missions such as, e.g., LUVOIR (\citealt{luvo19}) and HabEx (\citealt{gaud20}) could aid in the discovery of exomoons, albeit not being the missions' primary objective. The Habitable Worlds Observatory (HWO), a future mission proposed by the National Academies' Pathways to Discovery in Astronomy and Astrophysics for the 2020s (\citealt{HWO}), may also be beneficial to the discovery of exomoons as discussed in \citet{Limbach_2024}. The authors argue that HWO, through eclipse detection in exoplanet light curves, will be capable of detecting potential exomoons, especially those with a host planet orbital separation of $\sim$1-10 au.

\section*{Acknowledgements}
This work has been supported in part through computational resources provided by The Open Science Grid (OSG). In addition, S.D.P. and M.C. acknowledge support from the Department of Physics, University of Texas at Arlington.  B.Q. acknowledges support from the Department of Physics \& Astronomy, East Texas A\&M University ({\it formerly Texas A\&M University-Commerce}).

\section*{Data Availability}

The data underlying this article will be shared on reasonable request to the corresponding author.

\newpage


\bibliographystyle{mnras}
\bibliography{submoon}

\begin{thebibliography}{}
\makeatletter
\relax
\def\mn@urlcharsother{\let\do\@makeother \do\$\do\&\do\#\do\^\do\_\do\%\do\~}
\def\mn@doi{\begingroup\mn@urlcharsother \@ifnextchar [ {\mn@doi@} {\mn@doi@[]}}
\def\mn@doi@[#1]#2{\def\@tempa{#1}\ifx\@tempa\@empty \href {http://dx.doi.org/#2} {doi:#2}\else \href {http://dx.doi.org/#2} {#1}\fi \endgroup}
\def\mn@eprint#1#2{\mn@eprint@#1:#2::\@nil}
\def\mn@eprint@arXiv#1{\href {http://arxiv.org/abs/#1} {{\tt arXiv:#1}}}
\def\mn@eprint@dblp#1{\href {http://dblp.uni-trier.de/rec/bibtex/#1.xml} {dblp:#1}}
\def\mn@eprint@#1:#2:#3:#4\@nil{\def\@tempa {#1}\def\@tempb {#2}\def\@tempc {#3}\ifx \@tempc \@empty \let \@tempc \@tempb \let \@tempb \@tempa \fi \ifx \@tempb \@empty \def\@tempb {arXiv}\fi \@ifundefined {mn@eprint@\@tempb}{\@tempb:\@tempc}{\expandafter \expandafter \csname mn@eprint@\@tempb\endcsname \expandafter{\@tempc}}}

\bibitem[\protect\citeauthoryear{{Akeson} et~al.,}{{Akeson} et~al.}{2013}]{Akeson2013}
{Akeson} R.~L.,  et~al., 2013, \mn@doi [\pasp] {10.1086/672273}, \href {https://ui.adsabs.harvard.edu/abs/2013PASP..125..989A} {125, 989}

\bibitem[\protect\citeauthoryear{{Bonfanti}, {Ortolani}, {Piotto}  \& {Nascimbeni}}{{Bonfanti} et~al.}{2015}]{2015AA...575A..18B}
{Bonfanti} A.,  {Ortolani} S.,  {Piotto} G.,   {Nascimbeni} V.,  2015, \mn@doi [\aap] {10.1051/0004-6361/201424951}, \href {https://ui.adsabs.harvard.edu/abs/2015A&A...575A..18B} {575, A18}

\bibitem[\protect\citeauthoryear{{Cincotta} \& {Sim{\'o}}}{{Cincotta} \& {Sim{\'o}}}{1999}]{Cincotta1999}
{Cincotta} P.,  {Sim{\'o}} C.,  1999, \mn@doi [Celestial Mechanics and Dynamical Astronomy] {10.1023/A:1008355215603}, \href {https://ui.adsabs.harvard.edu/abs/1999CeMDA..73..195C} {73, 195}

\bibitem[\protect\citeauthoryear{{Cincotta} \& {Sim{\'o}}}{{Cincotta} \& {Sim{\'o}}}{2000}]{Cincotta2000}
{Cincotta} P.~M.,  {Sim{\'o}} C.,  2000, \mn@doi [\aaps] {10.1051/aas:2000108}, \href {https://ui.adsabs.harvard.edu/abs/2000A&AS..147..205C} {147, 205}

\bibitem[\protect\citeauthoryear{{Cincotta}, {Giordano}  \& {Sim{\'o}}}{{Cincotta} et~al.}{2003}]{Cincotta2003}
{Cincotta} P.~M.,  {Giordano} C.~M.,   {Sim{\'o}} C.,  2003, \mn@doi [Physica D Nonlinear Phenomena] {10.1016/S0167-2789(03)00103-9}, \href {https://ui.adsabs.harvard.edu/abs/2003PhyD..182..151C} {182, 151}

\bibitem[\protect\citeauthoryear{{Cuntz}, {Quarles}, {Eberle}  \& {Shukayr}}{{Cuntz} et~al.}{2013}]{Cuntz2013}
{Cuntz} M.,  {Quarles} B.,  {Eberle} J.,   {Shukayr} A.,  2013, \mn@doi [\pasa] {10.1017/pas.2013.011}, \href {https://ui.adsabs.harvard.edu/abs/2013PASA...30...33C} {30, e033}

\bibitem[\protect\citeauthoryear{{Dobos}, {Charnoz}, {P{\'a}l}, {Roque-Bernard}  \& {Szab{\'o}}}{{Dobos} et~al.}{2021}]{Dobos_2021}
{Dobos} V.,  {Charnoz} S.,  {P{\'a}l} A.,  {Roque-Bernard} A.,   {Szab{\'o}} G.~M.,  2021, \mn@doi [\pasp] {10.1088/1538-3873/abfe04}, \href {https://ui.adsabs.harvard.edu/abs/2021PASP..133i4401D} {133, 094401}

\bibitem[\protect\citeauthoryear{{Dombard}, {Cheng}, {McKinnon}  \& {Kay}}{{Dombard} et~al.}{2012}]{Dombard_2012}
{Dombard} A.~J.,  {Cheng} A.~F.,  {McKinnon} W.~B.,   {Kay} J.~P.,  2012, \mn@doi [Journal of Geophysical Research (Planets)] {10.1029/2011JE004010}, \href {https://ui.adsabs.harvard.edu/abs/2012JGRE..117.3002D} {117, E03002}

\bibitem[\protect\citeauthoryear{{Domingos}, {Winter}  \& {Yokoyama}}{{Domingos} et~al.}{2006}]{Domingos2006}
{Domingos} R.~C.,  {Winter} O.~C.,   {Yokoyama} T.,  2006, \mn@doi [\mnras] {10.1111/j.1365-2966.2006.11104.x}, \href {https://ui.adsabs.harvard.edu/abs/2006MNRAS.373.1227D} {373, 1227}

\bibitem[\protect\citeauthoryear{{Forgan}}{{Forgan}}{2018}]{forgan_2018}
{Forgan} D.,  2018, \mn@doi [Research Notes of the American Astronomical Society] {10.3847/2515-5172/aae8e6}, \href {https://ui.adsabs.harvard.edu/abs/2018RNAAS...2..191F} {2, 191}

\bibitem[\protect\citeauthoryear{Gaudi et~al.,}{Gaudi et~al.}{2020}]{gaud20}
Gaudi B.~S.,  et~al., 2020, The Habitable Exoplanet Observatory (HabEx) Mission Concept Study Final Report (\mn@eprint {arXiv} {2001.06683})

\bibitem[\protect\citeauthoryear{{Harada} et~al.,}{{Harada} et~al.}{2023}]{Harada_2023}
{Harada} C.~K.,  et~al., 2023, \mn@doi [\aj] {10.3847/1538-3881/ad011c}, \href {https://ui.adsabs.harvard.edu/abs/2023AJ....166..208H} {166, 208}

\bibitem[\protect\citeauthoryear{{Heller} \& {Barnes}}{{Heller} \& {Barnes}}{2013}]{2013AsBio..13...18H}
{Heller} R.,  {Barnes} R.,  2013, \mn@doi [Astrobiology] {10.1089/ast.2012.0859}, \href {https://ui.adsabs.harvard.edu/abs/2013AsBio..13...18H} {13, 18}

\bibitem[\protect\citeauthoryear{{Heller} \& {Hippke}}{{Heller} \& {Hippke}}{2024}]{Heller2024}
{Heller} R.,  {Hippke} M.,  2024, \mn@doi [Nature Astronomy] {10.1038/s41550-023-02148-w}, \href {https://ui.adsabs.harvard.edu/abs/2024NatAs...8..193H} {8, 193}

\bibitem[\protect\citeauthoryear{{Heller} et~al.,}{{Heller} et~al.}{2014}]{2014AsBio..14..798H}
{Heller} R.,  et~al., 2014, \mn@doi [Astrobiology] {10.1089/ast.2014.1147}, \href {https://ui.adsabs.harvard.edu/abs/2014AsBio..14..798H} {14, 798}

\bibitem[\protect\citeauthoryear{{Heller}, {Rodenbeck}  \& {Bruno}}{{Heller} et~al.}{2019}]{Heller2019}
{Heller} R.,  {Rodenbeck} K.,   {Bruno} G.,  2019, \mn@doi [\aap] {10.1051/0004-6361/201834913}, \href {https://ui.adsabs.harvard.edu/abs/2019A&A...624A..95H} {624, A95}

\bibitem[\protect\citeauthoryear{{Herman}, {Zhu}  \& {Wu}}{{Herman} et~al.}{2019}]{2019AJ....157..248H}
{Herman} M.~K.,  {Zhu} W.,   {Wu} Y.,  2019, \mn@doi [\aj] {10.3847/1538-3881/ab1f70}, \href {https://ui.adsabs.harvard.edu/abs/2019AJ....157..248H} {157, 248}

\bibitem[\protect\citeauthoryear{{Jagtap}, {Quarles}  \& {Cuntz}}{{Jagtap} et~al.}{2021}]{Jagtap_2021}
{Jagtap} O.,  {Quarles} B.,   {Cuntz} M.,  2021, \mn@doi [\pasa] {10.1017/pasa.2021.52}, \href {https://ui.adsabs.harvard.edu/abs/2021PASA...38...59J} {38, e059}

\bibitem[\protect\citeauthoryear{{Kawahara} \& {Masuda}}{{Kawahara} \& {Masuda}}{2019}]{2019AJ....157..218K}
{Kawahara} H.,  {Masuda} K.,  2019, \mn@doi [\aj] {10.3847/1538-3881/ab18ab}, \href {https://ui.adsabs.harvard.edu/abs/2019AJ....157..218K} {157, 218}

\bibitem[\protect\citeauthoryear{{Kipping} et~al.,}{{Kipping} et~al.}{2022}]{Kipping2022}
{Kipping} D.,  et~al., 2022, \mn@doi [Nature Astronomy] {10.1038/s41550-021-01539-1}, \href {https://ui.adsabs.harvard.edu/abs/2022NatAs...6..367K} {6, 367}

\bibitem[\protect\citeauthoryear{{Kipping} et~al.,}{{Kipping} et~al.}{2024}]{Kipping2024}
{Kipping} D.,  et~al., 2024, \mn@doi [arXiv e-prints] {10.48550/arXiv.2401.10333}, \href {https://ui.adsabs.harvard.edu/abs/2024arXiv240110333K} {p. arXiv:2401.10333}

\bibitem[\protect\citeauthoryear{Kollmeier \& Raymond}{Kollmeier \& Raymond}{2018}]{kollmeier_2018}
Kollmeier J.~A.,  Raymond S.~N.,  2018, \mn@doi [Monthly Notices of the Royal Astronomical Society: Letters] {10.1093/mnrasl/sly219}, 483, L80

\bibitem[\protect\citeauthoryear{{Kozai}}{{Kozai}}{1962}]{Kozai1962}
{Kozai} Y.,  1962, \mn@doi [\aj] {10.1086/108790}, \href {https://ui.adsabs.harvard.edu/abs/1962AJ.....67..591K} {67, 591}

\bibitem[\protect\citeauthoryear{{Kreidberg}, {Luger}  \& {Bedell}}{{Kreidberg} et~al.}{2019}]{Kreidberg2019}
{Kreidberg} L.,  {Luger} R.,   {Bedell} M.,  2019, \mn@doi [\apjl] {10.3847/2041-8213/ab20c8}, \href {https://ui.adsabs.harvard.edu/abs/2019ApJ...877L..15K} {877, L15}

\bibitem[\protect\citeauthoryear{{LUVOIR Team}}{{LUVOIR Team}}{2019}]{luvo19}
{LUVOIR Team} T.,  2019, The LUVOIR Mission Concept Study Final Report (\mn@eprint {arXiv} {1912.06219})

\bibitem[\protect\citeauthoryear{{Laskar} \& {Robutel}}{{Laskar} \& {Robutel}}{2001}]{saba}
{Laskar} J.,  {Robutel} P.,  2001, Celestial Mechanics and Dynamical Astronomy, \href {https://ui.adsabs.harvard.edu/abs/2001CeMDA..80...39L} {80, 39}

\bibitem[\protect\citeauthoryear{Levison, Walsh, Barr  \& Dones}{Levison et~al.}{2011}]{Levison_2011}
Levison H.,  Walsh K.,  Barr A.,   Dones L.,  2011, \mn@doi [Icarus] {https://doi.org/10.1016/j.icarus.2011.05.031}, 214, 773

\bibitem[\protect\citeauthoryear{{Lidov}}{{Lidov}}{1962}]{Lidov1962}
{Lidov} M.~L.,  1962, \mn@doi [\planss] {10.1016/0032-0633(62)90129-0}, \href {https://ui.adsabs.harvard.edu/abs/1962P&SS....9..719L} {9, 719}

\bibitem[\protect\citeauthoryear{{Limbach}, {Lustig-Yaeger}, {Vanderburg}, {Vos}, {Heller}  \& {Robinson}}{{Limbach} et~al.}{2024}]{Limbach_2024}
{Limbach} M.~A.,  {Lustig-Yaeger} J.,  {Vanderburg} A.,  {Vos} J.~M.,  {Heller} R.,   {Robinson} T.~D.,  2024, \mn@doi [\aj] {10.3847/1538-3881/ad4a75}, \href {https://ui.adsabs.harvard.edu/abs/2024AJ....168...57L} {168, 57}

\bibitem[\protect\citeauthoryear{{Moraes}, {Borderes-Motta}, {Winter}  \& {Monteiro}}{{Moraes} et~al.}{2022}]{moraes_2022}
{Moraes} R.~A.,  {Borderes-Motta} G.,  {Winter} O.~C.,   {Monteiro} J.,  2022, \mn@doi [\mnras] {10.1093/mnras/stab3576}, \href {https://ui.adsabs.harvard.edu/abs/2022MNRAS.510.2583M} {510, 2583}

\bibitem[\protect\citeauthoryear{{Moraes}, {Borderes-Motta}, {Winter}  \& {Mour{\~a}o}}{{Moraes} et~al.}{2023}]{moraes_2023}
{Moraes} R.~A.,  {Borderes-Motta} G.,  {Winter} O.~C.,   {Mour{\~a}o} D.~C.,  2023, \mn@doi [\mnras] {10.1093/mnras/stad314}, \href {https://ui.adsabs.harvard.edu/abs/2023MNRAS.520.2163M} {520, 2163}

\bibitem[\protect\citeauthoryear{Murray \& Dermott}{Murray \& Dermott}{2000}]{Murray_Dermott_2000}
Murray C.~D.,  Dermott S.~F.,  2000, Solar System Dynamics.
Cambridge University Press

\bibitem[\protect\citeauthoryear{{Naoz}}{{Naoz}}{2016}]{Naoz2016}
{Naoz} S.,  2016, \mn@doi [\araa] {10.1146/annurev-astro-081915-023315}, \href {https://ui.adsabs.harvard.edu/abs/2016ARA&A..54..441N} {54, 441}

\bibitem[\protect\citeauthoryear{{National Academies of Sciences}}{{National Academies of Sciences}}{2021}]{HWO}
{National Academies of Sciences} E.,  2021, {Pathways to Discovery in Astronomy and Astrophysics for the 2020s}, \mn@doi{10.17226/26141.
}

\bibitem[\protect\citeauthoryear{OpenStax}{OpenStax}{2022}]{openstax}
OpenStax 2022, Astronomy 2e by OpenStax, second edn.
XanEdu Publishing Inc

\bibitem[\protect\citeauthoryear{{Quarles}, {Gautham Bhaskar}  \& {Li}}{{Quarles} et~al.}{2024}]{quarleshkpq}
{Quarles} B.,  {Gautham Bhaskar} H.,   {Li} G.,  2024, \mn@doi [arXiv e-prints] {10.48550/arXiv.2407.13901}, \href {https://ui.adsabs.harvard.edu/abs/2024arXiv240713901Q} {p. arXiv:2407.13901}

\bibitem[\protect\citeauthoryear{{Rein} \& {Liu}}{{Rein} \& {Liu}}{2012}]{rebound}
{Rein} H.,  {Liu} S.~F.,  2012, \mn@doi [\aap] {10.1051/0004-6361/201118085}, \href {https://ui.adsabs.harvard.edu/abs/2012A&A...537A.128R} {537, A128}

\bibitem[\protect\citeauthoryear{{Rein} \& {Spiegel}}{{Rein} \& {Spiegel}}{2015}]{reboundias15}
{Rein} H.,  {Spiegel} D.~S.,  2015, \mn@doi [\mnras] {10.1093/mnras/stu2164}, \href {https://ui.adsabs.harvard.edu/abs/2015MNRAS.446.1424R} {446, 1424}

\bibitem[\protect\citeauthoryear{{Rein} \& {Tamayo}}{{Rein} \& {Tamayo}}{2015}]{reboundwhfast}
{Rein} H.,  {Tamayo} D.,  2015, \mn@doi [\mnras] {10.1093/mnras/stv1257}, \href {https://ui.adsabs.harvard.edu/abs/2015MNRAS.452..376R} {452, 376}

\bibitem[\protect\citeauthoryear{{Rein}, {Tamayo}  \& {Brown}}{{Rein} et~al.}{2019}]{reboundhighorder}
{Rein} H.,  {Tamayo} D.,   {Brown} G.,  2019, \mn@doi [\mnras] {10.1093/mnras/stz2503}, \href {https://ui.adsabs.harvard.edu/abs/2019MNRAS.489.4632R} {489, 4632}

\bibitem[\protect\citeauthoryear{{Rosario-Franco}, {Quarles}, {Musielak}  \& {Cuntz}}{{Rosario-Franco} et~al.}{2020}]{Rosario-Franco2020}
{Rosario-Franco} M.,  {Quarles} B.,  {Musielak} Z.~E.,   {Cuntz} M.,  2020, \mn@doi [\aj] {10.3847/1538-3881/ab89a7}, \href {https://ui.adsabs.harvard.edu/abs/2020AJ....159..260R} {159, 260}

\bibitem[\protect\citeauthoryear{{Stassun} et~al.,}{{Stassun} et~al.}{2019}]{Stassun2019}
{Stassun} K.~G.,  et~al., 2019, \mn@doi [\aj] {10.3847/1538-3881/ab3467}, \href {https://ui.adsabs.harvard.edu/abs/2019AJ....158..138S} {158, 138}

\bibitem[\protect\citeauthoryear{{Teachey} \& {Kipping}}{{Teachey} \& {Kipping}}{2018}]{Teachey2018}
{Teachey} A.,  {Kipping} D.~M.,  2018, \mn@doi [Science Advances] {10.1126/sciadv.aav1784}, \href {https://ui.adsabs.harvard.edu/abs/2018SciA....4.1784T} {4, eaav1784}

\bibitem[\protect\citeauthoryear{{Teachey}, {Kipping}, {Burke}, {Angus}  \& {Howard}}{{Teachey} et~al.}{2020}]{Teachey2020}
{Teachey} A.,  {Kipping} D.,  {Burke} C.~J.,  {Angus} R.,   {Howard} A.~W.,  2020, \mn@doi [\aj] {10.3847/1538-3881/ab7001}, \href {https://ui.adsabs.harvard.edu/abs/2020AJ....159..142T} {159, 142}

\bibitem[\protect\citeauthoryear{{Timmermann}, {Heller}, {Reiners}  \& {Zechmeister}}{{Timmermann} et~al.}{2020}]{2020A&A...635A..59T}
{Timmermann} A.,  {Heller} R.,  {Reiners} A.,   {Zechmeister} M.,  2020, \mn@doi [\aap] {10.1051/0004-6361/201937325}, \href {https://ui.adsabs.harvard.edu/abs/2020A&A...635A..59T} {635, A59}

\bibitem[\protect\citeauthoryear{{Tinney}, {Butler}, {Marcy}, {Jones}, {Penny}, {McCarthy}  \& {Carter}}{{Tinney} et~al.}{2002}]{2002ApJ...571..528T}
{Tinney} C.~G.,  {Butler} R.~P.,  {Marcy} G.~W.,  {Jones} H. R.~A.,  {Penny} A.~J.,  {McCarthy} C.,   {Carter} B.~D.,  2002, \mn@doi [\apj] {10.1086/339916}, \href {https://ui.adsabs.harvard.edu/abs/2002ApJ...571..528T} {571, 528}

\bibitem[\protect\citeauthoryear{{Williams}, {Kasting}  \& {Wade}}{{Williams} et~al.}{1997}]{1997Natur.385..234W}
{Williams} D.~M.,  {Kasting} J.~F.,   {Wade} R.~A.,  1997, \mn@doi [\nat] {10.1038/385234a0}, \href {https://ui.adsabs.harvard.edu/abs/1997Natur.385..234W} {385, 234}

\bibitem[\protect\citeauthoryear{{Wittenmyer} et~al.,}{{Wittenmyer} et~al.}{2020}]{2020MNRAS.492..377W}
{Wittenmyer} R.~A.,  et~al., 2020, \mn@doi [\mnras] {10.1093/mnras/stz3436}, \href {https://ui.adsabs.harvard.edu/abs/2020MNRAS.492..377W} {492, 377}

\bibitem[\protect\citeauthoryear{{Zollinger}, {Armstrong}  \& {Heller}}{{Zollinger} et~al.}{2017}]{2017MNRAS.472....8Z}
{Zollinger} R.~R.,  {Armstrong} J.~C.,   {Heller} R.,  2017, \mn@doi [\mnras] {10.1093/mnras/stx1861}, \href {https://ui.adsabs.harvard.edu/abs/2017MNRAS.472....8Z} {472, 8}

\bibitem[\protect\citeauthoryear{{von Zeipel}}{{von Zeipel}}{1910}]{vonZeipel1910}
{von Zeipel} H.,  1910, \mn@doi [Astronomische Nachrichten] {10.1002/asna.19091832202}, \href {https://ui.adsabs.harvard.edu/abs/1910AN....183..345V} {183, 345}

\makeatother
\end{thebibliography}




\appendix






\section{Coordinate Rotation Transforms} \label{sec:appendix}
In \texttt{rebound}, the transformation from orbital elements (e.g., inclination and ascending node) are performed essentially through a series of rotations \citep[see][]{Murray_Dermott_2000}, where the inclination is a rotation about the $x$-axis and the ascending node about the $z$-axis.  This is adequate for our 3-body simulations (star--planet--moon) because only one orbit is tilted/rotated relative to the reference plane.  For the 4-body simulations, the orbital plane of the submoon is measured relative to the tilted/rotated host moon's orbital plane, which is itself displaced relative to the central body (see Fig.~\ref{fig:orbit}).  

We show here how the initial Cartesian coordinates of the moon and submoon are rotated to obtain the correct inclination, along with the counter-rotation of the coordinates needed to return the submoon's coordinates relative to the moon. Parameters with the subscript "m" refer to the moon (i.e., natural satellite) and parameters with the subscript "sm" refer to the submoon.  The $\mathbf{r}$ designates vectors containing the Cartesian position coordinates of the bodies relative to their primaries $\{x,\ y,\ z\}$, whereas the $\mathbf{r}^\prime$ vectors are position coordinates relative to the host star as used for \texttt{rebound}.  The orbital inclination is a rotation using the matrix

\begin{align}
    R_x(\theta) \ = \ \begin{pmatrix}
1 & 0 & 0\\
0 & \cos{\theta} & -\sin{\theta} \\
0 & \sin{\theta} & \cos{\theta}
\end{pmatrix},
\end{align} 
where $\theta$ represents the rotation angle used ($i_{\rm m}$ or $i_{\rm sm}$)

To get the submoon's coordinates relative to the host star, we must first find the Cartesian vectors $\mathbf{r}$ of the planet $\mathbf{r}_{\rm p}$, moon $\mathbf{r}_{\rm m}$, and submoon $\mathbf{r}_{\rm sm}$ relative to their respective host bodies.  The rotation matrix for orbital inclination is first applied to the submoon to tilt its orbit by $i_{\rm sm}$, which is relative to the host moon's orbital plane.  Then, we add the host moon's Cartesian vector $\mathbf{r}_{\rm m}$ as a Galilean transformation.  The resulting vector must now be rotated by the host moon's inclination $i_{\rm m}$, which is relative to the host planet's orbital plane.  Finally, another Galilean transformation is employed using the host planet's Cartesian vector $\mathbf{r}_{\rm p}$.  As a result, we find the Cartesian vector of the submoon relative to the central body $\mathbf{r}^\prime_{\rm sm}$ as

\begin{align} \label{eqn:sm_rot}
R_{\rm x}(i_{\rm m})\mathbf{r}_{\rm m} + \mathbf{r}_{\rm p} &= \mathbf{r}^\prime_{\rm m}, \\
R_{\rm x}(i_{\rm m})R_{\rm x}(i_{\rm sm})\mathbf{r}_{\rm sm} + \mathbf{r}^\prime_{\rm m} &= \mathbf{r}^\prime_{\rm sm}, \quad\text{or}\\
R_{\rm x}(i_{\rm m})\left[R_{\rm x}(i_{\rm sm})\mathbf{r}_{\rm sm} + \mathbf{r}_{\rm m}\right] + \mathbf{r}_{\rm p} &= \mathbf{r}^\prime_{\rm sm}.
\end{align}

Another rotation $R_{\rm x}(i_{\rm p})$ would be applied to the left-hand side if we considered an inclined planet's orbital plane relative to some reference plane (e.g., ecliptic).  A similar process can be implemented to determine the velocity vector $\mathbf{v}^\prime$ for the submoon relative to the central body using velocity vectors $\mathbf{v}_{\rm m}$ and $\mathbf{v}_{\rm sm}$ relative to the host bodies of the moon and submoon, respectively.  

An inverse transformation is also needed to obtain the submoon's orbital inclination from the Cartesian vector of the submoon relative to the central body $\mathbf{r}^\prime_{\rm sm}$ to the respective vector relative to the host moon.  Starting from
Eq.~\ref{eqn:sm_rot}, we distribute and rearrange terms to obtain

\begin{align}
R_{\rm x}(i_{\rm m})R_{\rm x}(i_{\rm sm})\mathbf{r}_{\rm sm}  &= \mathbf{r}^\prime_{\rm sm} - \left[\mathbf{r}_{\rm p} - R_{\rm x}(i_{\rm m})\mathbf{r}_{\rm m}\right], \\
R_{\rm x}(i_{\rm m})R_{\rm x}(i_{\rm sm})\mathbf{r}_{\rm sm}  &= \mathbf{r}^\prime_{\rm sm} - \mathbf{r}^\prime_{\rm m}, 
\end{align}
and the inverse rotation $R_{\rm x}(-i_{\rm m})$ is applied using the host moon's current orbital inclination $i_{\rm m}$ to get

\begin{align}
R_{\rm x}(i_{\rm sm})\mathbf{r}_{\rm sm}  &= R_{\rm x}(-i_{\rm m})\left[\mathbf{r}^\prime_{\rm sm} - \mathbf{r}^\prime_{\rm m}\right], 
\end{align}
where the primed Cartesian vectors represent positions relative to the central body.  To get the submoon's inclination relative to the orbital plane of the host moon, an inverse rotation $R_{\rm x}(-i_{\rm m})$ must be applied to the difference vector $\left[\mathbf{r}^\prime_{\rm sm} - \mathbf{r}^\prime_{\rm m}\right]$.  Note that $\mathbf{r}_{\rm sm}$ describes the submoon's Cartesian vector relative to the host moon.









\bsp	
\label{lastpage}
\end{document}